\documentclass[%
 reprint,
 amsmath,amssymb,
 aps,
]{revtex4-2}

\usepackage{graphicx}
\usepackage{dcolumn}
\usepackage{bm}

\begin{document}

\preprint{APS/123-QED}

\title{Hybrid coupling optomechanically assisted nonreciprocal photon blockade }

\author{Yong-Pan Gao}%
\email{gaoyongpan@bupt.edu.cn}
 \affiliation{School of Electronic Engineering and the State Key Laboratory of Information Photonics and Optical Communications, Beijing University of Posts and Telecommunications, Beijing 100876, China}

\author{Chuan Wang}%
 \email{wangchuan@bnu.edu.cn}
\affiliation{%
 School of Artificial Intelligence, Beijing Normal University, Beijing 100875, China
}%

\date{\today}

\begin{abstract}
The properties of open quantum system in quantum information science is now extensively investigated more generally by the research community as a fundamental issue for a variety of applications. Usually, the states of the open quantum system might be disturbed by the decoherence which will reduce the fidelity in the quantum information processing. So it is better to eliminate the influence of the environment. However, as part of the composite system, rational use of the environment system could be beneficial to quantum information processing. Here we theoretically studied the environment induced quantum nonlinearity and energy spectrum tuning method in the optomechanical system. And we found that the dissipation coupling of the hybrid dissipation and dispersion optomechanical system can induce the coupling between the environment and system in the cross-Kerr interaction form. When the symmetry is broken with a directional pumping environment, the system exhibits the non-reciprocal behavior during the photon excitation and photon blockade for the clockwise and counterclockwise modes of the whispering-gallery mode microcavity. Furthermore, we believe that the cross-Kerr coupling can also be used in a more widely region in quantum information processing and quantum simulation.
\end{abstract}

\maketitle


\section{Introduction}

Quantum information processing and quantum metrology rely on the efficient manipulation of the qubits which have attracted intense interest during the past decades. It has been found that the system for quantum information processing are not isolated and may be characterized by the open quantum system. Furthermore, the decoherence of the environment may induce noise and reduce the fidelity of the quantum states. On the other hand, the rational use of the environment system may be beneficial to the quantum system. Here we are intending to investigate the dynamics of the dissipation process based on the cavity optomechanical system. Recently, cavity optomechanical system is introduced into the study of quantum information science. Cavity optomechanical system \cite{bowen2015quantum,aspelmeyer2014cavity,RevModPhys.86.1391,Kippenberg:07,kippenberg2008cavity,bahl2013brillouin} is an artificial microstructure that combines the interaction between the mechanical mode and the optical mode with various effect could be observed on it. For example, the optomechanical induced transparency \cite{weis2010optomechanically,kronwald2013optomechanically,jing2015optomechanically,zhu2020optomechanically,lu2018optomechanically,lu2017optomechanically,PhysRevA.88.013804,Dong:14,safavi2011electromagnetically,PhysRevA.81.041803}, optomechanical assisted nonlinearity \cite{PhysRevA.100.063827,roque2020nonlinear,PhysRevLett.114.253601,PhysRevLett.114.013601,PhysRevE.98.032201,PhysRevLett.119.153901,miri2018optomechanical,cao2016tunable,gao2016effective,gao2017optomechanically}, phonon lasers \cite{PhysRevLett.112.143602,wang2018polarization,PhysRevApplied.8.044020,PhysRevLett.113.053604,PhysRevApplied.10.064037,zhang2018phase}, in the semi-classical region; the photon blockade \cite{PhysRevLett.107.063601,PhysRevLett.107.063602,PhysRevA.88.023853,li2019nonreciprocal,PhysRevA.99.043818,PhysRevA.98.013826,PhysRevLett.121.153601},  mechanical resonator cooling \cite{PhysRevLett.110.153606,vanner2013cooling,PhysRevA.85.051803,PhysRevA.98.023860,PhysRevA.84.053838,PhysRevLett.124.103602},  nonclassical state preparation \cite{PhysRevLett.116.163602,PhysRevA.93.033853,PhysRevA.94.053807,PhysRevA.101.033812,PhysRevLett.114.093602,PhysRevLett.112.080503,PhysRevLett.117.143601,PhysRevLett.110.010504,PhysRevLett.123.113601} , and the fundamental quantum properties \cite{PhysRevD.98.022003,PhysRevA.96.043824,PhysRevX.8.011031,PhysRevA.93.022510,PhysRevA.100.062516} in the full quantum region.

 The geometric deformation characteristics of the optomechanics and the geometric deformation correlation of its optical properties endow the optical machine with unique and excellent physical properties. Different from other systems, such as the cavity-atom coupled system \cite{RevModPhys.73.565,RevModPhys.87.1379,mucke2010electromagnetically,bao2012efficient,PhysRevLett.114.023601} and cavity-magnetic coupled system \cite{PhysRevLett.117.123605,PhysRevLett.116.223601,PhysRevA.100.043831,PhysRevB.101.054412,PhysRevLett.120.057202}, optomechanics provides us a platform to investigate the properties of the non-static coupling between the system and the environment. The dissipative optomechanics describes the coupling of the optical mode to the environment of a cavity \cite{PhysRevLett.103.223901,weiss2013strong,hryciw2015tuning,PhysRevLett.116.233604,kronwald2014dissipative,PhysRevA.95.023844,PhysRevLett.107.213604,PhysRevLett.112.076402}, when it is combined with the dispersion optomechanics \cite{weis2010optomechanically,jing2015optomechanically,zhu2020optomechanically,lu2018optomechanically}, the hybridization of the dissipative optomechanics with the dispersion optomechanics shows potential applications in pure optical manipulation. The hybrid systems have become a key strategy in microwave electromechanical system \cite{PhysRevLett.102.207209,PhysRevA.97.063820,regal2008measuring,PhysRevLett.101.197203,asano2018opto}. Recently, various studies have shown that such kind of optomechanical system with both dispersive coupling and dissipative coupling can be realized in the whispering-gallery cavities \cite{PhysRevX.4.021052,vahala2003optical,PhysRevLett.103.053901,collot1993very,lin2014barium,ryu2004high}.

 In this study, we present a full quantum approach that describes the dissipation and dispersion hybrid coupled optomechanical system. First, we developed the self-Kerr effect based on the dispersive coupling, and illustrated the cross-Kerr effect of the environment caused by dissipative coupling on the system. In order to show the potential applications of the cross Kerr effect, relying on the directional coherent manipulation of the cross-Kerr interaction,  this approach is further applied to achieve the tunable non-reciprocity between the intrinsic optical clockwise and counter-clockwise modes, and also the non-reciprocal photon blockade.

\section{The dissipation-dispersion hybrid optomechanics}

\begin{figure}
	\centering
	\includegraphics[width=0.8\linewidth]{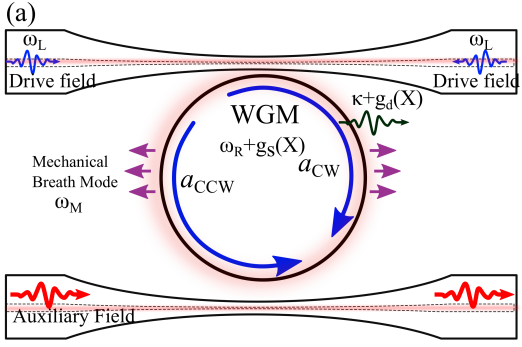}
	\caption{The hybrid coupling optomecanics and its energy level scheme. The  whisper gallery mode (WGM)  has an unstable frequency $\omega_{R}+g_\text{s}(X)$  and dissipation $\kappa+g_\text{d}(X)$. Both the clockwise (CW) and counterclockwise (CCW) of thie WGM can be excited. An auxiliary field is weakly coupled with the CCW mode and tuning it resonant features through the dissipation induced cross-kerr effect. Here we studied the quantum feature of the weak drive field which is strong coupled with the cavity with the frequency $\omega_{\text{L}}$.  }
	\label{fig:optomechanics}
\end{figure}

The system is shown in FIG.\ref{fig:optomechanics}, the whispering-gallery mode (WGM) microcavity supports both the optical mode and the mechanical breath mode. Due to the optomechanical interaction, the mechanical breach mode will change both the curvature \cite{PhysRevApplied.12.014060} and effective length of the microcavity, in which the dissipation and dispersion hybrid coupling optomechanics appears. The mechanical mode has the resonant frequency of $\omega_M$ and negligible damping rate, while the optical mode has the position dependent frequency $\omega_R + g_s(X)$ and dissipation rate $\kappa + g_d (X)$. The optomechanical microcavity is coupled with the add-drop fiber tapers. For the two fiber waveguides, the add fiber is strongly coupled with the cavity, and the weak driving field is input and output with the frequency $\omega_L$ through this fiber; the drop fiber is weakly coupled with the cavity (The feed back of the cavity is negligible), the input field plays the role of controllable environment which is defined as the auxiliary field.
The Hamiltonian of the dissipation-dispersion hybrid optomechanical system in FIG.\ref{fig:optomechanics} could be expressed as (with $\hbar=1$)
\begin{equation}
\hat{H}=\omega_{\text{R}} \hat{a}^{\dagger} \hat{a}+\omega_{\text{M}} \hat{c}^{\dagger} \hat{c}+\sum_{q} \omega_{q} \hat{b}_{q}^{\dagger} \hat{b}_{q}+\hat{H}_{\text{d}}+\hat{H}_{\text{s}}+\hat{H}_{\gamma}
\end{equation}

Here the optical mode has the frequency of $\omega_R$ with the annihilation operator denoted as $\hat{a}$. $\hat{c}$ denotes the annihilation operator of the mechanical mode whose frequency is $\omega_M$. $\hat{b}_q$ is the annihilation operator of the environment with the frequency $\omega_q$. And $H_d$ corresponds to the dissipation coupling Hamiltonian, $H_s$ represents the dispersion coupling Hamiltonian, and $H_{\gamma}$ shows the dissipation part of the system. The present system could be achieved both in the microwave electro-mechanical system \cite{PhysRevLett.102.207209}  and optical WGM system. The microwave system offer easy manipulation and processing, yet have limited thermal noise, while the present system enlarge the hybrid optomechics from the microwave to optical wavelength. It can have much less thermal noise than the microwave case, but is usually hard to approach the strong coupling regime of the system because the co-existence of the dissipation and the dispersion coupling.

The dissipative optomechanical coupling part could be described as
\begin{equation}
\hat{H}_{\text{d}}=g_\text{d}(X)\equiv X\frac{\text{d}\,\kappa}{\text{d} x}(\hat{b}_q\hat{a}^\dagger-\hat{b}_q^\dagger\hat{a} ).
\end{equation}
here $X=\hat{c}^\dagger+\hat{c}$, and the dispersion coupling term is
\begin{equation}
\hat{H}_{\text{s}}=g_\text{s}(X)\hat{a}^\dagger\hat{a}=g_{\text{s}}\hat{X}\hat{a}^\dagger\hat{a}.
\end{equation}

For simplicity, we set $\hat{A}^\dagger=\sum_{q}\hat{b}_q\hat{a}^\dagger$. And the Hamilton could be written as
\begin{equation}
\hat{H}=\frac{\omega_{\text{R}}}{\hat{n}_q} \hat{a}^{\dagger} \hat{a}+\omega_{\text{M}} \hat{c}^{\dagger} \hat{c}+X[ig_{\text{d}}(\hat{A^\dagger-\hat{A}})+\frac{g_{\text{s}}}{\hat{n}_q}\hat{A}^\dagger\hat{A}].
\end{equation}

As $\hat{n}_q=\sum_q\hat{b}_{q}^\dagger\hat{b}_q$, we assignment the above Hamilton as
\begin{subequations}
	\begin{align}
	\hat{H}_0&=\Omega_\text{R}\hat{A}^\dagger\hat{A}+\omega_\text{M}\hat{c}^\dagger\hat{c},\\
	\hat{W}&=X\left[i g_{\text{d}}(\hat{A}^\dagger-\hat{A})+\frac{g_{\text{s}}}{\hat{n}_{q}}\hat{A}^\dagger\hat{A}\right],\\
	\hat{H}&=\hat{H}_0+\hat{W}.	
	\end{align}	
\end{subequations}
By choosing the quadrature amplitudes as $\hat{P}=i(\hat{A}^\dagger-\hat{A})$ and $\hat{Q}=\hat{A}^\dagger+\hat{A}$, then we have
\begin{equation}
\hat{W}=X [g_{\text{d}}\hat{P}+\frac{g_{\text{s}}}{4|B|^2}(\hat{Q}^2+\hat{P^2})].
\end{equation}

\begin{figure}
	\centering
	\includegraphics[width=0.8\linewidth]{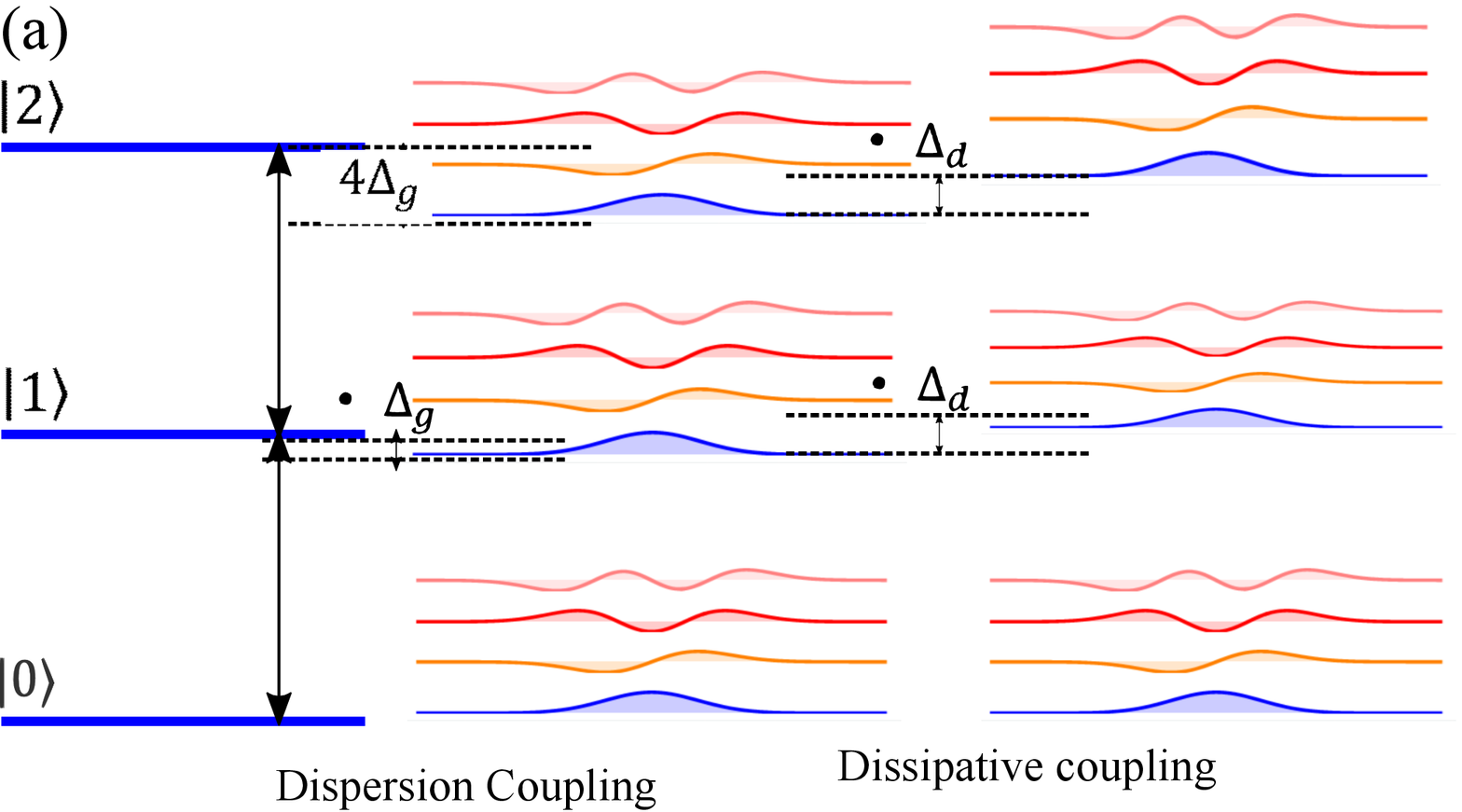}
	\includegraphics[width=0.8\linewidth]{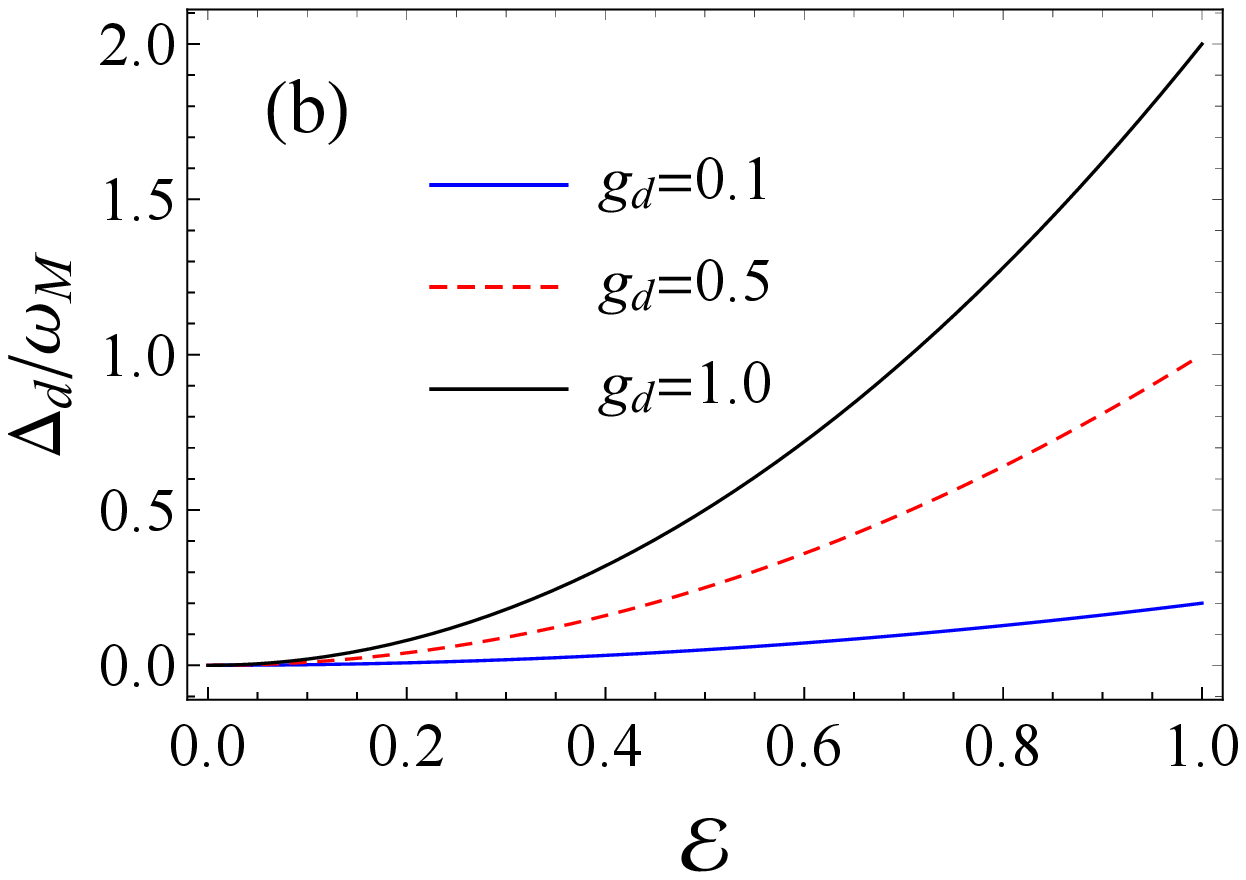}
	\caption{The energy scheme and shift of the hybrid optomechanical system. (a). How the dispersion and dissipation coupling modulation the energy scheme. (b). The energy shift $\Delta_d$ various with the auxiliary field strength $\mathcal{E}$ under different dissipation coupling strength $g_\text{d}/\omega_{\text{M}}=0.1,0.5,1$.}
	\label{fig:optomechanics-level}
\end{figure}

With the definition as $B=\langle\sum_q \hat{b}_q\rangle$, we perform the transformation by using the operator  $U_1=\text{e}^{S_1}$,where $S_1=i\frac{g_{\text{d}}|B|^2}{g_{\text{s}}}(\hat{A}^\dagger+\hat{A})$. Then we can get
\begin{equation}
\begin{split}
\hat{H}=&\frac{\Omega_R}{\hat{n}_{p}}(A-i\frac{g_{\text{d}}\hat{n}_q}{g_{\text{s}}})(A^\dagger+i\frac{g_{\text{d}}\hat{n}_q}{g_{\text{s}}})+\omega_\text{M}\hat{c}^\dagger\hat{c}\\
+&X\left[\frac{g_{\text{s}}}{\hat{n}}\hat{A}^\dagger\hat{A}+\frac{g_{\text{d}}^2\hat{n}_q}{g_{\text{s}}}\right],
\end{split}
\end{equation}
here the term $\hat{A}-\hat{A}^\dagger$  is the environment driven term that can be eliminated by the out-phase drive. Then, the Hamiltonian of the system could be changed to
\begin{equation}
\hat{H}=\frac{\Omega_R}{\hat{n}}\hat{A}^\dagger\hat{A}+\omega_M\hat{c}^\dagger\hat{c}+X\left[\frac{g_{\text{s}}}{\hat{n}}\hat{A}^\dagger\hat{A}+\frac{g_{\text{d}}^2\hat{n}_q}{g_{\text{s}}}\right].
\end{equation}

The environment is single mode $\hat{n}_q=\hat{b}^\dagger\hat{b}$. Then by using the transformation operator $S_2=\exp{(\hat{c}^\dagger-\hat{c})(\frac{g_{\text{s}}}{\hat{n}}\hat{A}^\dagger\hat{A}+\frac{g_{\text{d}}^2\hat{n}}{g_{\text{s}}})}$, the system could be changed to
\begin{equation}
\hat{H}=\left(\omega_\text{R}-\frac{g_{\text{s}}^2}{\omega_\text{M}}\hat{a}^\dagger\hat{a}-\frac{2g_{\text{d}}^2}{\omega_\text{M}}\hat{b}^\dagger\hat{b}\right) \hat{a}^\dagger\hat{a}+\omega_{\text{M}} \hat{c}^{\dagger} \hat{c}.\label{eq:cross}
\end{equation}
Previously, the cross-Kerr effect is usually discussed in the semi-classical region which induced directly \cite{PhysRevA.93.023844,PhysRevA.91.043822,zhang2017effects,PhysRevA.100.052306}. For example, the cross-Kerr effect between the mechanical mode and optical mode \cite{PhysRevA.99.043837,PhysRevLett.112.203603}. Here we  consider the cross-Kerr nonlinearity as a port which is open to the environment, and provides us a tool for the control of the cavity.

\section{Energy structure and its modulation with the auxiliary field}
As the environment plays the role of the cross-Kerr term, in order to study the dynamics of the system, here an auxiliary filed is introduced to simulate the environment which is stronger than the driving field. The environment for the directional features of the CW and CCW fields are different for the field in the waveguide.

The pump strength of the auxiliary field on the cavity is set as  $\mathcal{E}$, then frequency shift of  the CCW mode could be denoted as $\frac{2g_{\text{d}}^2\mathcal{E}^2}{\omega_\text{M}}\langle \alpha|\hat{b}^\dagger\hat{b}|\alpha\rangle$ according to Eq.\ref{eq:cross}. $|\alpha\rangle$ represents the coherent state. Here it is obvious that the frequency shift is different in the CW and CCW directions. We assume the intensity of the  probe beam is weaker compared with the pumping beam. And the probe beam is bidirectional, the pumping beam is only coupled with the CW mode. Then the Hamiltonian of the two modes could be expressed as
\begin{subequations}\label{eq:bidirec}
	\begin{align}
\hat{H}_{\text{CW}}&=\left(\omega_\text{R}-\frac{g_{\text{s}}^2}{\omega_\text{M}}\hat{a}_{\text{CW}}^\dagger\hat{a}_{\text{CW}}-\frac{2g_{\text{d}}^2}{\omega_\text{M}}\mathcal{E}^2\right) \hat{a}_{\text{CW}}^\dagger\hat{a}_{\text{CW}}
+\omega_{\text{M}} \hat{c}^{\dagger} \hat{c},\\
\hat{H}_{\text{CCW}}&=\left(\omega_\text{R}-\frac{g_{\text{s}}^2}{\omega_\text{M}}\hat{a}_{\text{CCW}}^\dagger\hat{a}_{\text{CCW}}\right) \hat{a}_{\text{CCW}}^\dagger\hat{a}_{\text{CCW}}+\omega_{\text{M}} \hat{c}^{\dagger} \hat{c}.
	\end{align}
	\end{subequations}
From Eq.\ref{eq:bidirec}, we can find frequency shift of both the CW and CCW modes are related to the intensity of the auxiliary field.
And the eigen-energy of the two modes could be solved as
\begin{subequations}
	\begin{align}
	E_{n,\text{CW}}&=n\hbar\omega_R-\frac{g_{\text{s}}^2}{\omega_{\text{M}}}n^2-\frac{2g_{\text{d}}^2}{\omega_{\text{M}}}\mathcal{E}^2+n_{\text{M}}\omega_{M},\\
		E_{n,\text{CCW}}&=n\hbar\omega_R-\frac{g_{\text{s}}^2}{\omega_{\text{M}}}n^2+n_{\text{M}}\omega_{\text{M}}.
	\end{align}
\end{subequations}

As shown in FIG.\ref{fig:optomechanics-level}, the energy-level structure generation of our scheme is presented. Except that the energy level distribution is no longer uniform,  the mechanical system will generate many mechanical sub-level which is different from the pure Kerr case. Thus, the dispersion coupling will shift the energy level with different amplitude and there are several mechanical dressed energy level generated. Then the energy could be shifted by the dissipation coupling generation, and the variance is related to the auxiliary field strength. In FIG.\ref{fig:optomechanics-level} (b) we show the frequency shift varies with the auxiliary field strength $\mathcal{E}$ under the dissipation coupling strength of $g_{\text{d}}=0.1,0.5,1$.

\section{The photon excitation and photon blockade of the hybrid coupling system}

As the energy gap is no longer equally spaced and auxiliary field is adjustable, the  non-reciprocal photon blockade could be achieved in our hybrid coupling system. The dynamical equation of the system without environment noise is
\begin{equation}\label{eq:dyn}
\dot{\hat{a}}(t)=(i\Delta_i-\kappa)\hat{a}_{i}(t)+e^{-i P(t)}\mathcal{E}_p+\mathcal{O}\left(\mathcal{E}_p^{2}a_{\mathrm{in}}\right).
\end{equation}
Here, $i$ denotes the mode CW or CCW. The detunings are $\Delta_{\text{CW}}=\Delta-\frac{g_{der}^2}{\omega_M}-\frac{2g_{dis}^2}{\omega_M}\mathcal{E}^2$,  $\Delta_{\text{CCW}}=\Delta-\frac{g_{der}^2}{\omega_M}$, and $\Delta=\omega_\text{L}-\omega_\text{R}$ is the detuning of the laser from the bare cavity frequency.  $P(t)=e^{-i H_{m} t} P e^{i H_{m} t}$,  $ H_{m}=\omega_{M}\hat{c}^\dagger\hat{c}$ is the free mechanical Hamilton.  Here we assume the pumping beam is the environment which is not affected by the feedback. By integrating Eq.\ref{eq:dyn} under the weak driven condition, we find the spectrum function of the  system could be solved as \cite{PhysRevLett.107.063601,PhysRevLett.107.063602,makri1995tensor,vagov2011real,gradshteyn2014table}:
\begin{equation}
S\left(\Delta\right) \simeq \kappa \sum_{n=-\infty}^{\infty} A_{n} \frac{\kappa_{n}}{\kappa_{n}^{2}+\left(\Delta_i-n \omega_{m}\right)^{2}},
\end{equation}
where $A_{n}=e^{-\eta^{2}(2 N+1)} I_{n}[2 \eta^{2} \sqrt{N(N+1)}]((N+1) / N)^{n / 2}$.  $N$ denotes the total photon number, in which $N(\omega)=1 /\left(e^{\hbar \omega / k_{B} T}-1\right)$ for the thermal state, $k_B$ is the Boltzmann constant.

 \begin{figure}
	\centering
	\includegraphics[width=0.8\linewidth]{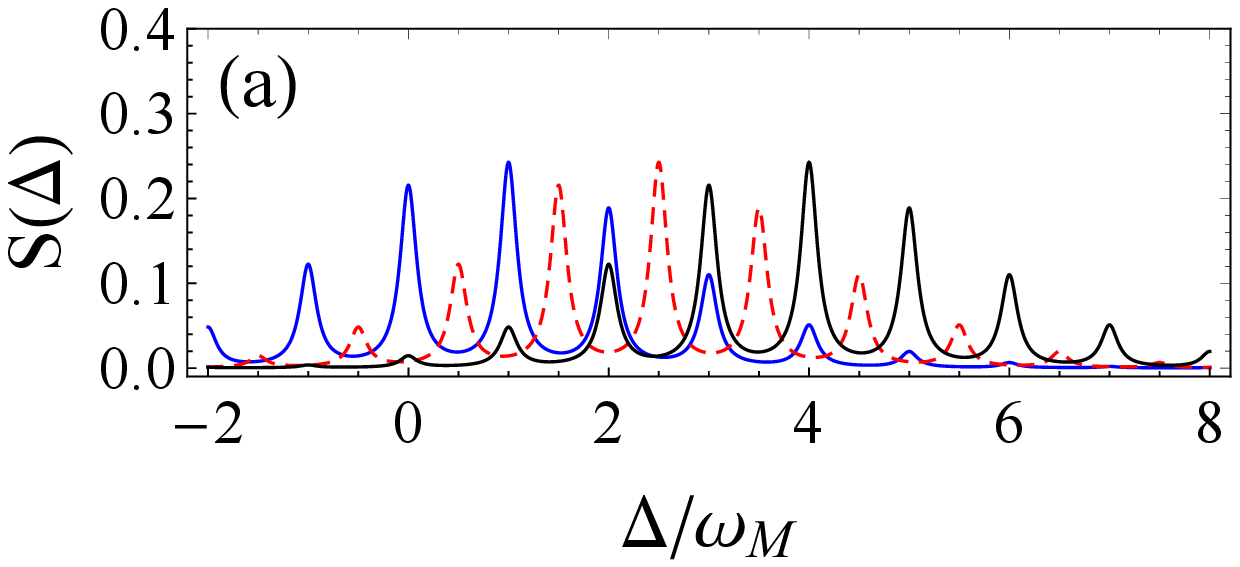}
	\includegraphics[width=0.8\linewidth]{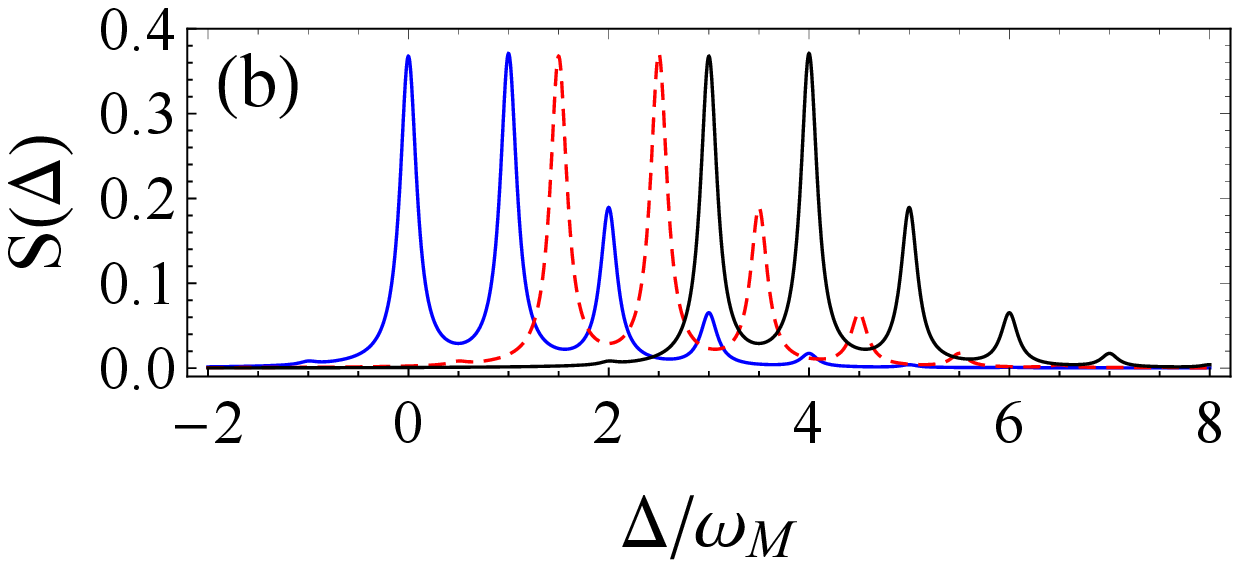}
	\caption{The photon number shift with high thermal phonon number and lower photon number. (a). The photon excitation spectrum various with  frequency detuning $\Delta$ under 1 thermal phonon environment. (b). The photon excitation spectrum various with  frequency detuning $\Delta$ under 0.01 thermal phonon environment. In both figure, we have $\kappa/\omega_{\text{M}}=0.1$, the quality factor of the mechanical mode is $10^4$, and the dispersion coupling strength $g_{\text{s}}/\omega_{M}=1$. The shift strength is $\Delta_{\text{d}}/\omega_{\text{M}}=0.5$ for the red line, while it is 1 for the black line.}
	\label{fig:photon-shift}
\end{figure}

To further clarify the properties of the hybrid coupling system, we show the photon excitation spectrum in FIG.\ref{fig:photon-shift}. Here the black line represents the photon excitation of CW mode, while the red and blue line show the auxiliary field shift photon excitation spectrum of the CCW mode. In figure FIG.\ref{fig:photon-shift}(a), the thermal photon number is set as 1. We can find that there are resonance peaks at both the positive and negative detuning regions. When the thermal photon number is chosen as 0.01 in FIG.\ref{fig:photon-shift} (b), there are only resonance peaks in the positive detuning region. It is because the thermal noise will bridge different mechanical sub-levels.

One more important thing is that the auxiliary field will cause energy level displacement. Here we show the cross-Kerr induced shift $\Delta_d=0$, $\Delta_d=0.5 \omega_m$ and  $\Delta_d=\omega_m$ with different line in this figure, respectively. There are obviously spectrum shift for both cases. However, the reciprocity of the excitation spectrum is different in FIG.\ref{fig:photon-shift}(a) and (b). When the thermal photon number is higher (1 thermal photon in the environment), the blue line and the blackline is partial reciprocition when $\Delta/\eta= 1$, or $2$, while it is fully reciprocal under the lower thermal environment condition. Moreover, we can also achieve the fully reciprocity by tuning the cross-Kerr induced frequency shift.  when compare the blue and red line, we can find their peak points are alternately distributed. Then we can also achieve the fully reciprocition by choosing suitable $\Delta$.

Beyond the energy levels of the system, we also studied the properties of the correlation functions. The dynamical equation of  the second order term is
\begin{equation}\label{eq:co}
\dot{\hat{a}}^2(t)=2(i\Delta_i+i\Delta_g-\kappa)\hat{a}^2(t)+\mathcal{E}_pe^{-iP(t)}a(t).
 \end{equation}


By integrating the above equation, $a^2$ could be solved. Then, the second order correction function $g^{(2)}_i(0)=\langle {a^\dagger}^2a^2\rangle/\langle a^\dagger a\rangle^2$ can be written as
\begin{equation}
\begin{split}
	g_i^{(2)}(0)&= \operatorname{Re} \sum_{n, m, p} \frac{B_{n, m, p}}{\left[\kappa+i\left(\Delta_i-n \omega_{M}\right)\right]\left[\kappa-i\left(\Delta_i-m \omega_{M}\right)\right]} \\
	 &\times \frac{2 \kappa^{3}}{\left[2 \kappa-i\left(2 \Delta_i+2 \Delta_{g}-p \omega_{m}\right)\right]S^2\left(\Delta\right)}.\label{eq:corre}
	 \end{split}
\end{equation}
 For simplify, we set the temperature is near zero, then we have  $B_{n, m, p}=$ $e^{-2 \eta^{2}}\left(\eta^{2}\right)^{p} W_{n, p}(\eta) W_{m, p}(\eta) / n ! m ! p ! $, $W_{n, p}(\eta)=(-1)^{n} U\left[-n, 1-n+p, \eta^{2}\right]$. The $U[a, b, x]$ is a confluent hypergeometric function,  and $\eta=g_{der}/\omega_M$ is a dimensionless parameter.

  \begin{figure}
 	\centering
 	\includegraphics[width=0.8\linewidth]{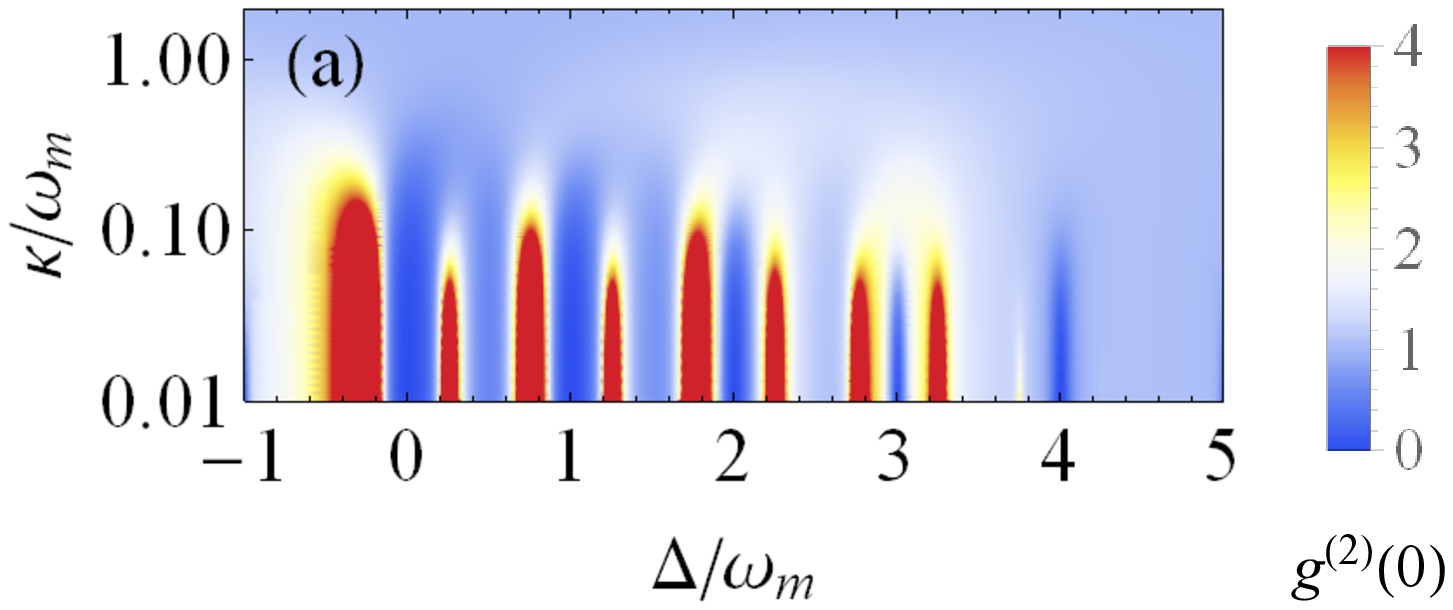}
 	\includegraphics[width=0.8\linewidth]{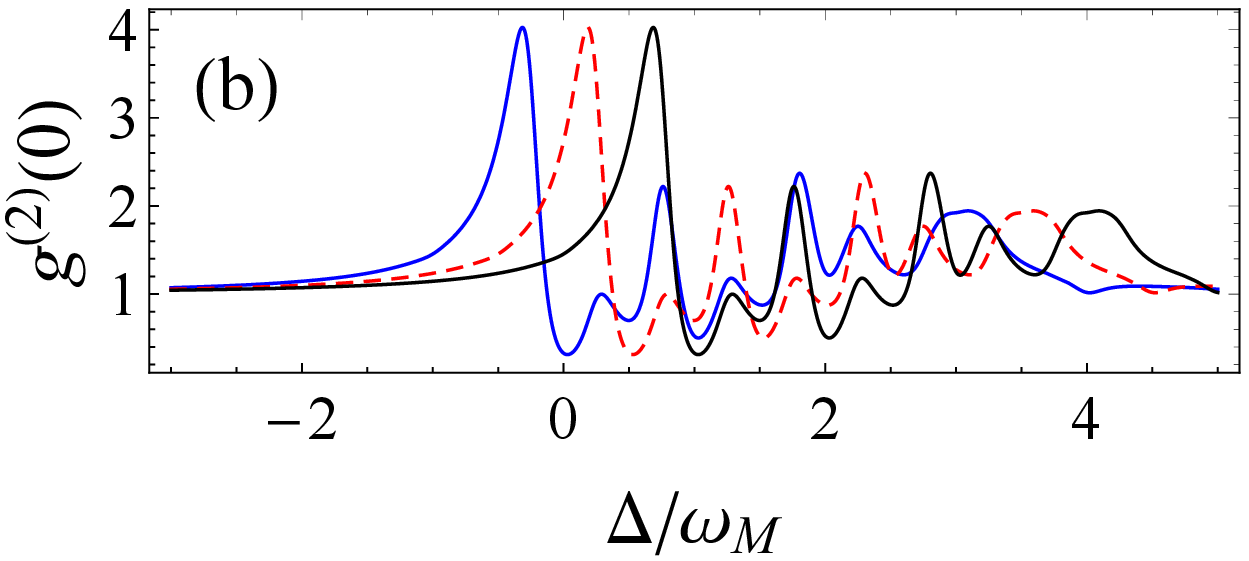}
 	\includegraphics[width=0.8\linewidth]{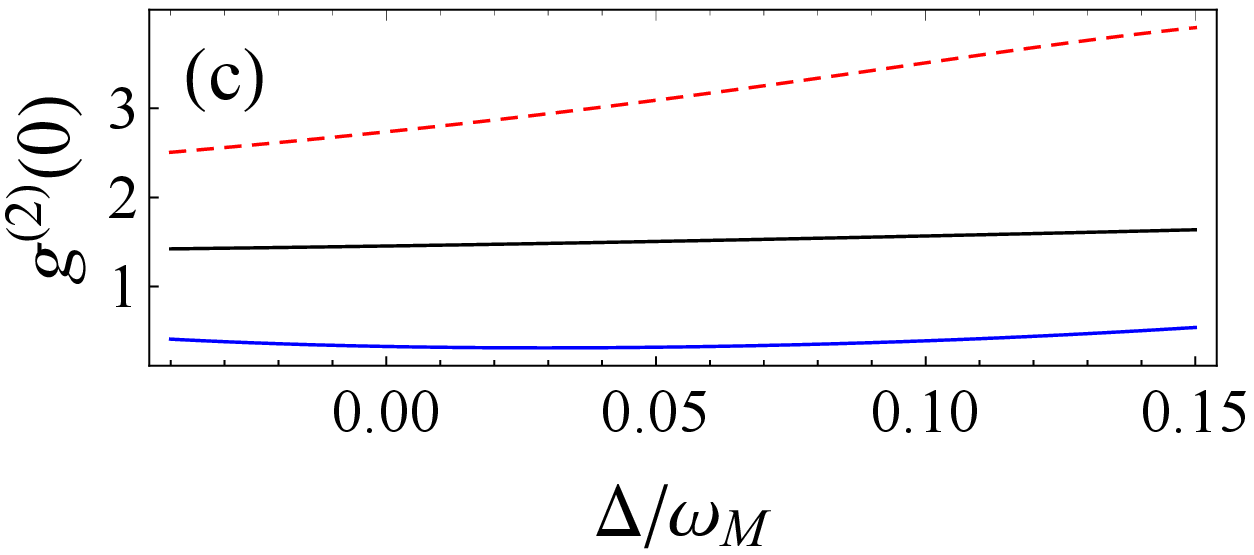}
 	\caption{The second correlation function and its shift of the hybrid optomechanics. (a). The second correlation function with 0.001 thermal phonons. (b). The correlation spectrum shift with the auxiliary field. The $\kappa/\omega_{M}=0.15$ and thermal phonons number is 0.001. We ignore the loss of mechanical mode. The shift strength is $\Delta_{\text{d}}/\omega_{\text{M}}=0.5$ for the red line, while it is 1 for the black line. (c). The zoom of (b).}
 	\label{fig:den-g2}
 \end{figure}

 According to the second-order correlation function given in Eq.\ref{eq:corre}, the photon blockade characteristic of the hybrid optomechanical system is also studied.  In FIG.\ref{fig:den-g2} (a), we report the theoretical calculations of the CW (black line) and CCW (red and blue line) correlation spectrum under different driving strength and cavity dissipation. We find that the second-order correlation function is almost zero  when the detuning $\Delta/\omega_M$ is $0,1,2,3,4$ which displays the photon blockade, and it decreases along with the decrement of the parameter $\kappa$. So, we choose $\kappa=0.15$ as a key parameter in our study. Then the coherent environment can shift the $g^{(2)}(0)$ of the CCW mode in FIG.\ref{fig:den-g2} (b). Thus, the non-reciprocity could be achieved by choosing the parameter for the $g^{(2)}(0)$, we can easily achieve non-reciprocity. In order to show the non-reciprocity more obviously, we show its zoom in  FIG.\ref{fig:den-g2} (c). However, the degree of non-reciprocity under different pumping strength is different. When the shift is $0.5\omega_{M}$, the $g^{(2)}(0)$ is ranging between 1 and 2. When the shift is $\omega_{M}$, the $g^{(2)}(0)$ is above 2. So, when we give a frequency as  $\omega_{M}$, the system shows better performance of the non-reciprocity.

The non-reciprocal photon blockade is a part of the non-reciprocity feature of the second-order correlation function. To quantitatively study the nonreciprocal feature of the this system, we define the photon nonreciprocity as
\begin{equation}
\mathcal{R}(\Delta)=\frac{\left(\log(g_{CW}^{(2)})-\log(g_{CCW}^{(2)})\right)^2}{|\log(g_{CW}^{(2)})|+|\log(g_{CCW}^{(2)})|}.
\label{eq:nonre}
\end{equation}
There, the denominator represents the sum of the correlation function of the CW and CCW mode, while the numerator is the size of their difference. We choose the logarithm operation of the correlation as the 1 is the decision point of photon correlation.

\begin{figure}
	\centering
	\includegraphics[width=0.8\linewidth]{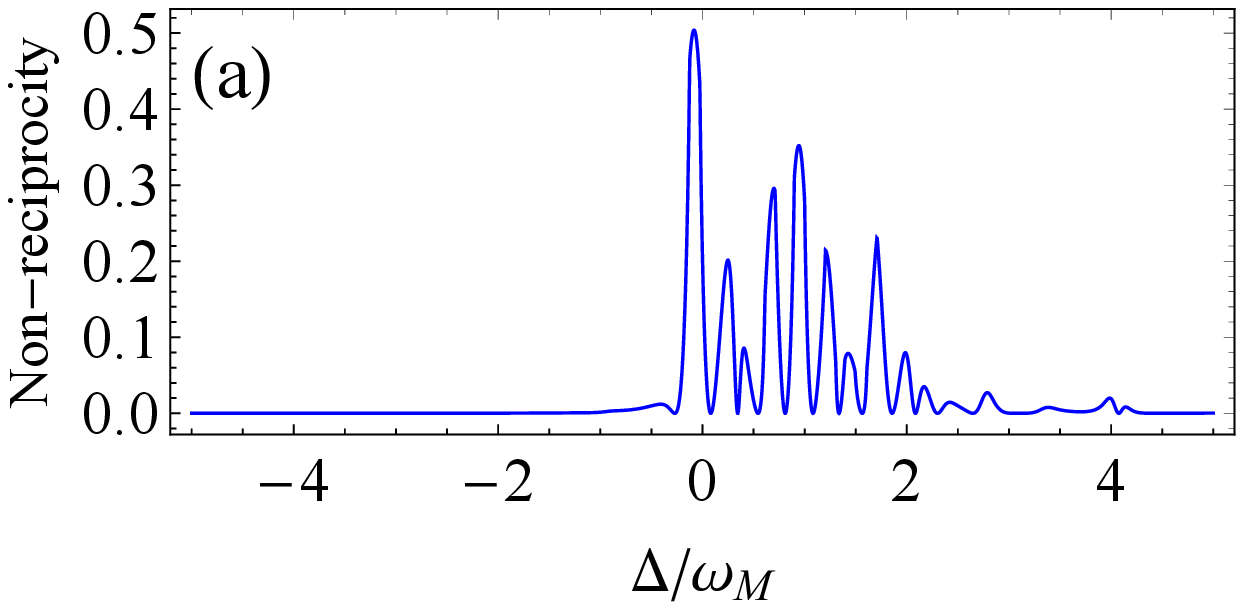}
	\includegraphics[width=0.8\linewidth]{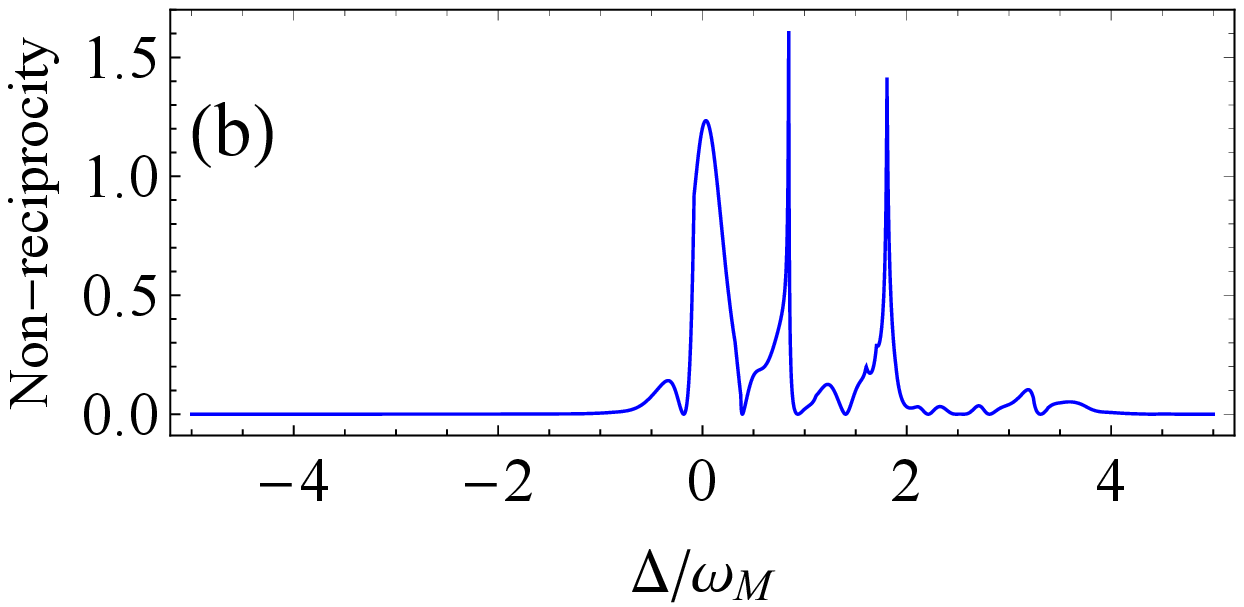}
	\includegraphics[width=0.8\linewidth]{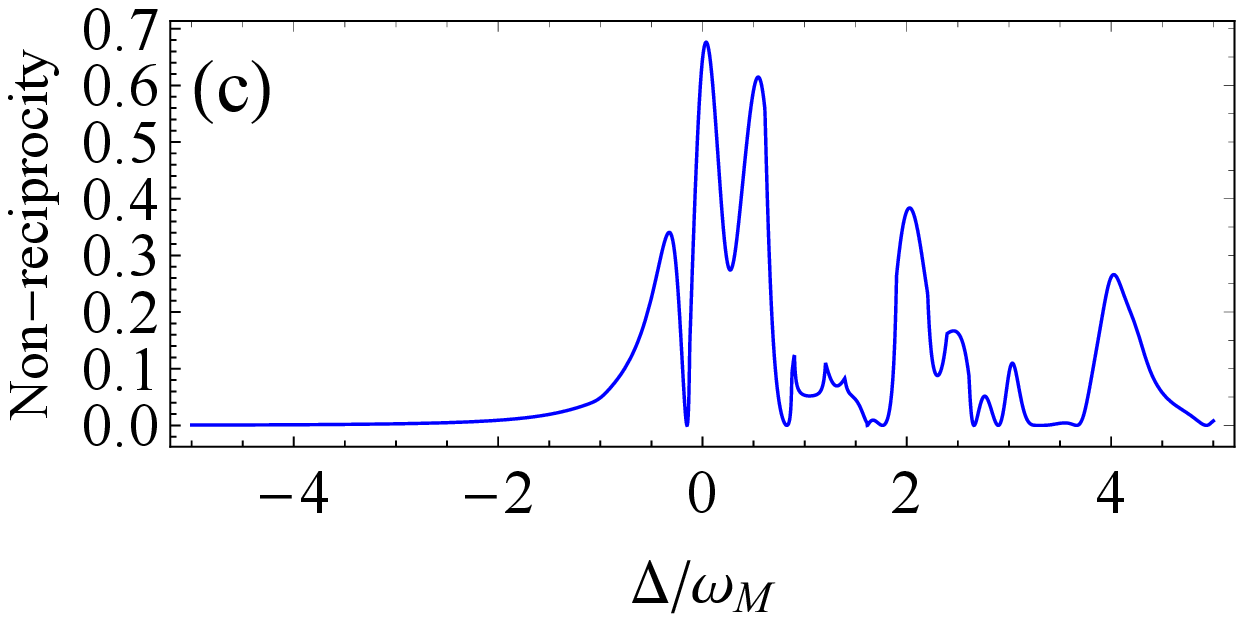}
	\caption{The Non-Reciprocity under different environment pumping. (a). The shift is $\Delta_\text{d}=0.1$. (b). The shift is $\Delta_\text{d}=0.5$. (c). The shift is $\Delta_\text{d}=1$.  In all figures, we have $\kappa/\omega_{\text{M}}=0.15$, the dispersion coupling strength $g_{\text{s}}/\omega_{M}=0.5$, the thermal phonon numbers is 0.001,and the mechanical dissipation is neglected.}
	\label{fig:nor-re1}
\end{figure}
The performance of  Eq.\ref{eq:nonre} is shown in FIG.\ref{fig:nor-re1}.  The coherent induced frequency shift is 0.1, 0.5, and 1 for the CCW mode, respectively. When the detuning is small as 0.1 as shown in  FIG.\ref{fig:nor-re1}(a), the main peak appears in the near-zero region, while the maximal value is smaller than $\log_{10}(4)$. So the non-reciprocal photon blockade will not occur. While when we set the environment induced shift as 0.5, as shown in FIG.\ref{fig:nor-re1}(b), there are transmission peaks in the resonance position ($\Delta/\omega_{\text{M}}=0,1,2$). It is because the resonance dip in the transmission spectrum of the CCW mode is shifted to the peak of the CW mode, and the non-reciprocal photon blockade occur. When we continuously increasing the shift to 1, as shown in FIG.\ref{fig:nor-re1}(c), the maximal values of the non-reciprocity decreases, for the resonant peaks of the CW and CCW modes meet again. When we return to FIG.\ref{fig:den-g2}(b)(c) again, we can find the non-reciprocity is not large as the Eq.\ref{eq:nonre}(b), but the non-reciprocal photon blockade can still occur.

\section{Summary}

In summary, we present a full quantum approach to study the dynamical behavior of the dissipation and dispersion coupled optomechanical system. The coupling would introduce the Kerr type nonlinearity, in particular, dissipative coupling will cause cross Kerr coupling between the cavity mode and the environment.  On the other hand, by exploiting the cross-Kerr nonlinearity in the system, the non-reciprocity of the second-order correlation function on the CW and CCW propagation optical modes is illustrated. Moreover, we find the photon blockade effect on both the CW and the CCW propagation directions. Our research give an insight view of the environment of the whispering gallery mode in the full quantum region. When we simply apply this scheme to an coherent environment, is shows excellent photon non-reciprocal blocking performance which will be the powerful tool in the  realization of  quantum networking, nonlinear quantum gate, quantum information processing.

\section{Acknowledgements}

The authors gratefully acknowledge the support from the Fundamental Research Funds for the Central Universities(BNU) and the National Natural Science Foundation of China through Grants Nos. 61622103.

\bibliography{bref}

\end{document}